# Effect of atmosphere and sintering time on the microstructure and mechanical properties at high temperatures of $\alpha$-SiC sintered with liquid phase $Y_2O_3$-$Al_2O_3$

**M. Castillo-Rodríguez, A. Muñoz, A. Domínguez-Rodríguez***

Departamento de Física de la Materia Condensada. Universidad de Sevilla.

Apartado 1065. 41080-Sevilla. Spain

The influence that the atmosphere ($N_2$ or Ar) and sintering time have on microstructure evolution in liquid-phase-sintered $\alpha$-SiC (LPS-$\alpha$-SiC) and on its mechanical properties at high temperature was investigated. The microstructure of the samples sintered in $N_2$ was equiaxed with a grain size of 0.70 µm and a density of 98% of the theoretical value regardless of the sintering time. In contrast, samples sintered in Ar had an elongated-grain microstructure with a density decreasing from 99% to 95% and a grain size increasing from 0.64 to 1.61 µm as the sintering time increased from 1 to 7 hours. The mechanical behaviour at 1450 °C showed the samples sintered in nitrogen to be brittle and fail at very low strains, with a fracture stress increasing from 400 to 800 MPa as the sintering time increased. In contrast, the samples sintered in Ar were quasi-ductile with increasing strain to failure as the sintering time increased, and a fracture stress strongly linked to the form and size of the grains. These differences in the mechanical properties of the two materials are discussed in the text. During mechanical tests, a loss of intergranular phase takes place in a region, between 50 and 150 µm thick, close to the surface of the samples — the effect being more important in the samples sintered in Ar.

________

* Corresponding author. Tel.: +34-95-455-7849; fax: +34-95-461-2097.

**1. Introduction**

In the last few decades, much effort has been put into developing structural ceramics due to the demand of industry for materials to be used in ever severer conditions. Amongst them, SiC has been one of the most widely studied due to its remarkable mechanical properties at high as well as at low temperatures, its low density, and its excellent resistance to corrosion and thermal shock. However, it is difficult to obtain fully dense SiC materials due to the covalent bonding between Si and C and the low diffusion coefficient of the two species. It has been necessary to develop sintering techniques to reduce temperatures and stresses. In this regard, Prochazka[1] in 1975, using B and C as additives, obtained a high densification of $\beta$-SiC powders at temperatures above 2000 °C. Coppola et al.[2] reported the same results with $\alpha$-SiC powders, and more recently several workers[3-8] have found that both $\beta$-SiC and $\alpha$-SiC powders can be sintered at temperatures lower than 2000 °C using liquid phase aids such as $Al_2O_3$ and $Y_2O_3$, and oxynitrates.

The mechanical properties of these liquid phase sintered SiC (LPS-SiC) materials depend strongly on the microstructure, and this in turn on the amount and type of the liquid phase, type of starting powder, temperature, time, and sintering atmosphere. Knowledge of the relationships between the sintering process parameters, the microstructure, and the mechanical properties is fundamental to obtaining SiC with desired properties.

Several workers have connected the microstructure of liquid-phase-sintered SiC to its mechanical properties at high temperatures. Nagaro et al.[9] studied the influence of the processing atmosphere ($N_2$ or Ar) on the constant strain rate compressive superplastic behaviour of LPS-$\beta$-SiC nanocrystals using 5.1%-wt $Al_2O_3$ and 3.8%-wt $Y_2O_3$ as additives. They concluded that the deformation is grain boundary sliding (GBS) with a decreasing strain prior to fracture for the samples sintered in $N_2$. Ortiz et al.[10] performed compressive tests at a constant strain rate in air at 1400 °C on LPS-$\beta$-SiC with 20%-vol of $Y_3Al_5O_{12}$ (YAG) using two sintering atmospheres ($N_2$ and Ar). They found that $N_2$-LPS SiC has a higher deformation resistance and higher ultimate compressive strength, but lower strain to failure, than Ar-LPS SiC.

Schneider et al.[11] studied the creep behaviour of a SiC composed of 10%-wt of $\alpha$-SiC and 90%-wt of $\beta$-SiC as starting powders, and sintered using $Y_2O_3$ and AlN as additives in different ratios. The samples presented most creep resistance when the additive ratio was 2:3 molar and when the samples were annealed in $N_2$ atmosphere after sintering. Jou et al.[12] studied the creep of an Ar-LPS $\alpha$-SiC, the liquid phase being $Al_2OC$ (10%-wt) and deformed at temperatures between 1550 and 1660 °C. They concluded that the deformation is controlled simultaneously by bulk diffusion processes and dislocation activity. In the same line, Gallardo-López et al.[13] studied the creep behaviour of an LPS-$\alpha$-SiC with a small amount of liquid phase — less than 2%-vol $Y_2O_3$ and $Al_2O_3$ — and deformed at temperatures between 1575 and 1700 °C. The deformation is due to GBS controlled simultaneously by volume diffusion of point defects and dislocation activity.

The objective of the present work was to correlate the microstructure and mechanical behaviour of an LPS-α-SiC, with $Y_2O_3$ and $Al_2O_3$ as liquid phase (10%-wt), and sintered in two atmosphere ($N_2$ and Ar) for different process times. The mechanical tests were performed at a constant strain rate, and the microstructure before and after deformation was determined by scanning electron microscopy (SEM).

## 2. Experimental techniques

The powder used in this work was composed of 90% α-SiC and 10% $Y_2O_3$ and $Al_2O_3$ as intergranular phase in the ratio 3/5 to get yttrium aluminium garnet $Y_3Al_5O_{12}$ (YAG). The powders were sintered at 1950 °C in two different atmospheres (Ar and $N_2$) for several times from 1 to 7 hours in the Department of Metallurgy and Material Science Engineering, Institute of Materials Science, University of Connecticut, USA. For more details see references[7,14].

The microstructure characteristics — grain size and morphology, pore distribution, and damage after high temperature tests — were determined using a Philips Model XL-30 scanning electron microscope (SEM; electron microscopy service, University of Seville, Spain) operating at 30 kV using backscattered and secondary electrons. Prior to observation, the samples were polished with diamond paste of grain size down to 1 µm, and plasma ($CF_4$ and $O_2$ in the ratio 1:6) etched for 2 hours (Plasma Asher K1050X, Emitech). In some cases, the samples were sputter-coated with gold to avoid charge effects during observation. The morphological characteristics were determined using a Videoplan image analyzer (Videoplan MOP 30, Kontron Electronik).

Specimens of 5×2.5×2.5 mm of the different materials were machined out and polished to a 3 µm finish. Uniaxial compression tests were performed in air at 1450 °C using an Instron universal testing machine (Model 1185) with the square face (2.5×2.5 mm) normal to the loading axis. The specimens were sandwiched between two solid-state-sintered SiC platelets in order to avoid damage to the $Al_2O_3$ pushing rods. A constant crosshead of 5 µm min$^{-1}$ was used, corresponding to an initial strain rate ($\dot{\varepsilon}$) of 1.7 10$^{-5}$ s$^{-1}$. The load-displacement data was recorded continuously and analyzed in terms of engineering stress and engineering strain.

## 3. Results and discussion

### 3.1. Microstructure characterization

Figure 1 is a typical SEM micrograph of the samples used in the present work. One observes that the SiC grains (dark contrast) are surrounded by the intergranular phase (light contrast). Also, the intergranular phase is homogeneously distributed, and there exists a small amount of residual porosity. This porosity is consistent with the density of the samples as measured by Archimedes' method. Table 1 lists the values of the density compared with the theoretical value of 3.310 ± 0.002 g/cm³. One observes that, consistent with the observed increase in porosity (Fig. 1), the density of Ar-LPS-α-SiC decreased from 99% to 95% of the theoretical value as the sintering time increased from 1 to 7 hours. On the contrary, the density did not vary with sintering time for the $N_2$-LPS-α-SiC samples, which had a value close to 98% of the theoretical density.

Figure 2 shows SEM micrographs of an Ar-LPS α-SiC. The intergranular phases (YAG) are located at the triple points, although a thin layer of glassy phase, unobserved under SEM, surrounds the SiC grains. The grain size *d* increased from 0.64 to 1.61 µm and the aspect ratio of the grains $F_{asp}$ also increased as the sintering time was prolonged from 1 to 7 hours (for more details, see Table 2). For the longest sintering time samples, both the grain size and the aspect ratio presented a bimodal distribution.

In Fig. 2, one observes the core-shell substructure within the SiC grains. This is indicative of grain growth occurring by the smaller grains dissolving in the glassy phase, followed by precipitation of C and Si atoms onto the largest grains. The core (dark contrast inside the grains) represents the original growing grains, and the shell represents the newly deposited material[15]. The difference in contrast is because the shell contains traces of Y, Al, and O that come from the glassy phase. This substructure is more difficult to distinguish in the largest grains because the relative smallness of the original grain compared to the final grain makes it less likely to be present in the cross section.

The anisotropic evolution of a few grains for the longer sintering time (7 hours) can be understood as the result of the presence of a certain amount of $\beta$ phase seeds in the $\alpha$-SiC starting powders. At the sintering temperature, the $\beta \rightarrow \alpha$ phase transformation takes place within these grains, resulting in grains with a composite $\beta/\alpha$ nature. The silicon and carbon species dissolve in the liquid and re-precipitate as $\alpha$-SiC onto $\alpha$-SiC regions in order to minimize the free energy of the system, thereby leading to the grains having a much higher aspect ratio which increases with annealing time[15]. The aspect ratio of the highly elongated grains hinders intergranular phase redistribution, and consequently the porosity of the samples sintered in Ar atmosphere increases with sintering time. For the 7-hour annealed samples, the main porosity is located near the largest grains (Fig. 2d).

Figure 3 shows the microstructure of the samples sintered in $N_2$ atmosphere for different sintering times. As in the samples sintered in Ar, the intergranular phases are located at the triple points of the SiC grains. The grains in all the samples sintered in $N_2$ have a nearly equiaxed form. During sintering in an $N_2$ atmosphere, nitrogen dissolves into the liquid-stage intergranular phase as well as into the SiC-grains. The presence of $N_2$ within the grains helps to stabilize the β polytype, thereby completely suppressing the β→α phase transformation responsible for the anisotropic growth of the β phase seeds in the α-SiC starting powders[16]. With respect to grain size, for this type of samples no appreciable change was observed with increasing sintering time, the value remaining constant and close to 0.70 μm. This inhibition of grain growth would be associated with the presence of $N_2$ in the intergranular phase, giving rise to increased solid-liquid interface energies and viscosity of the liquid phase[17, 18]. These effects would retard the solution-precipitation process and the diffusivity of the C and Si atoms through the liquid phase necessary for the growth of the grain.

The morphological parameters of the grains for the different types of sample were determined from SEM micrographs of different parts of the samples, using at least 350 grains in each case. Figure 4 shows histograms of the grain size distribution (equivalent planar diameter, $d=(4 \cdot area/\pi)^{1/2}$) of four of the materials studied. The grain distribution of the samples sintered for 7 hours in an Ar atmosphere is broader and shifted to larger grain sizes compared to the 1-hour case. The distribution is bimodal due to the existence of large elongated grains. As was noted above, the $N_2$-atmosphere sintered samples had almost equiaxed grains whose size remained unchanged by longer sintering times.

The other grain parameters measured for the different types of sample were the form factor (F=4π·area/perimeter$^2$) and the aspect ratio $F_{asp}$, defined as the ratio between the maximum length and the minimum width taken perpendicular to the length in the planar section of the grains. Table 2 gives the means and standard deviations of the morphological parameters, with all the parameters assumed to be normally distributed. One can observe in the table the dependence of the microstructure characteristics of the samples on the atmosphere and the process time.

*3.2. Mechanical tests*

Uniaxial compression tests were performed at constant cross head speed in air at 1450 °C. Figure 5a shows the engineering stress-engineering strain curves obtained for samples sintered in argon. The samples sintered from 1 to 5 hours presented very similar behaviour, with a yield stress between 350 and 450 MPa, a maximum stress between 500 and 550 MPa, and a strain prior to failure close to 8%. For the 7-hour sintered samples, the behaviour was different, with a yield stress of 650 MPa, a maximum stress close to 800 MPa, and a total strain of 11%.

Figure 5b shows the engineering stress-engineering strain curves obtained for samples sintered in nitrogen. In this case, there was a gradual increase of the yield stress (from 350 to 750 MPa) and of the maximum stress (from 400 MPa to 800 MPa) with sintering time. In all cases, the strain prior to failure was very small, being 2% in the best case.

The SEM micrographs of the deformed samples show that the intergranular phase was absent in a region (between 50 and 150 μm thick) close to the surface, (Fig.

6). This is a consequence of the movement of the intergranular phase towards the surface due to the squeezing of the glassy phase during the high temperature compression tests, together with oxidation processes, not as yet clearly identified[11, 19, 20], which take place in the surface of the sample in contact with the air. The absence of the intergranular phase close to the surface was less significant in the samples sintered in a nitrogen atmosphere than in those sintered in argon. In all cases, the absence of this intergranular phase gave rise to many pores, which increased in number with closeness to the surface.

The samples presented a viscous-looking surface layer (Fig. 7a) as a consequence of the oxidation processes that take place during deformation in air at 1450 °C. The EDS microanalysis of this layer showed the presence of Y, Al, and O (Fig. 7b). This layer may be formed because $Y_2O_3$ and $Al_2O_3$ constitute with $SiO_2$ a ternary system with a eutectic point below 1400 °C[13]. The presence of $SiO_2$ could be a consequence of surface oxidation of SiC during the compression test in air or of the reaction between SiC and $Al_2O_3$[13].

The SEM micrographs in the centre of the samples showed, in all the cases, that the size and form factor of the SiC grains remained unchanged after deformation. The deformation was thus not due to the transport of point defects (Nabarro-Herring or Coble mechanisms), because in such cases the grains would undergo almost the same deformation as the sample. This can be understood in terms of the low diffusion coefficient of Si and C at these temperatures (1450 °C)[21-24] and of the time used in the deformation tests (2-3 hours). The observed microstructure morphology indicates that grain boundary sliding is the main deformation mechanism in this system.

Although it is possible for dislocations to be activated under the conditions of the tests[25], it seems that, due to the high strain rate used in the tests ($2 \times 10^{-5}$ s$^{-1}$), this was not the accommodation process for the grain boundary sliding. The cavities and small cracks observed along the intergranular phase indicate that there were no effective accommodation processes, and that coalescence of the small cracks at the end of the deformation was responsible for the catastrophic failure of the samples. In the samples sintered in argon, the micrographs showed a high density of points at which there was significant loss of coherence of the grains (Figs. 8a and 8b), this being more evident in the sample sintered for 7 hours. The large aspect ratio grains that developed during the longer sintering time were responsible for hindering the interconnection of the cracks, thereby explaining the high deformation prior to failure in these samples. Moreover, grain boundary sliding was also more difficult, hence the higher flow stress.

The situation was different for the nitrogen atmosphere sintered samples. As was noted above, while the morphology of the grains remained unchanged with sintering time, the mechanical behaviour was markedly different (Fig. 5). During sintering, only small amounts of nitrogen incorporate into the SiC grains due to its low solubility in SiC (~100 ppm)[18]. In the intergranular glass phase, however, the solubility is much greater, reaching concentrations (depending of the sintering time) of for example 0.8% for 5 hours sintering[15]. Related studies on the incorporation of nitrogen into the intergranular glass phase have found significant increases in the high temperature viscosities[17, 18]. The replacement of twofold-coordinated oxygen by threefold-coordinated nitrogen in the glass increases the overall rigidity of the glass network, and, indeed, internal friction measurements in these nitrogen atmosphere sintered materials have shown that the incorporation of this element is the cause of the increase in viscosity at high temperatures[16].

The rigidity of the glass network and the higher binding energy due to the incorporation of nitrogen into the glass phase seem to be the causes of the increase in flow stress and maximum stress with annealing time. The microstructure of this type of material after deformation (Fig. 8c) shows the cracks to have coalesced, and long cracks can be seen. These cracks propagated through the intergranular phase, surrounding the SiC grains. The equiaxed form and the small size of the grains present in these materials is the reason for the catastrophic failure after deformation of only a few percent. The argon sintered materials are less brittle.

4. Conclusions

• The microstructure characterization of the as-received samples showed a homogeneous distribution of the intergranular phases, and a density close to 98% of the theoretical value independently of the sintering time for samples sintered in nitrogen. For samples sintered in argon, the density decreased as the annealing time increased, being 95% for the longest annealing time.

• The microstructure of the samples sintered in nitrogen was unchanged with increasing sintering time, and the grains were equiaxed with an average size of 0.70 μm. For the samples sintered in argon, the grain size increased from 0.64 to 1.61 μm, and the grains became ever more elongated as the sintering time increased from 1 to 7 hours.

- For the nitrogen sintered samples, the flow stress increased from 350 to 750 MPa and the maximum stress from 400 to 800 MPa as the annealing time was prolonged from 1 to 7 hours. This was a consequence of the incorporation of nitrogen into the intergranular phase.
- For the argon sintered samples, the flow stress remained between 350 and 400 MPa and the failure stress between 500 and 550 Mpa, almost independently of the sintering time. For the longest sintering time, the flow stress was 650 MPa and the failure stress was 800 MPa, due to the large elongated grains present in the microstructure.
- The almost unchanged morphology of the grains after deformation and the redistribution of the intergranular phase indicated that grain boundary sliding and the squeezing of the intergranular phase constituted the mechanism controlling deformation. The absence of any accommodation process led to the creation of cavities and cracks, and their coalescence was the cause of the failure of the material.
- The differences in behaviour between the nitrogen and the argon sintered samples were due to the different microstructures generated in the two types of sample, and to the incorporation of nitrogen into the intergranular phase leading to increased viscosity at high temperatures and rigidity of the glass network.
- The squeezing of the glass phase of a zone close to the surface, between 50 and 150 µm thick, increased the density of pores. This pore density was greater the closer to the surface. The phenomenon was more significant in the argon sintered samples due to the lower viscosity of the glassy phase. After deformation at 1450 °C, a viscous-looking layer appeared in the surface of the samples rich in Y, Al, and O.

**Acknowledgements**

The authors are grateful to Professors F. Guiberteau and A. L. Ortiz of the University of Extremadura for the preparation of the materials used in the work. The authors also thank the Ministry of Science and Technology for financial support through the project MAT2001-0799.

**References**

1. Prochazka, S., Sintering of silicon carbide, Proceedings of the Conference on Ceramics of High-Performance Applications (Hyannis, MA, 1973). Edited by Burke J. J., Gorum, A. E. and Katz R. M., Brook Hill, Chestnut Hill, MA, 1975, 239-252.

2. Coppola, J. A., Hawler H. A. and McMurtry, C. H., US Patent 4,123,286, 1978.

3. Omari M. and Takei, Pressureless sintering of SiC, J. Am. Ceram. Soc., 1982, 65, C-92.

4. Cutler R. A. and Jackson T. B., Liquid phase sintered silicon carbide, In Ceramic Materials and Components for Engines: Proceedings of the Third International Symposium. Edited by Tennery V. J., American Ceramic Society, Westerville, OH, 1989, 309-318.

5. Mulla M. A. and Krstic, V. D., Low-temperature pressureless sintering of $\beta$-silicon carbide with aluminium oxide and yttrium oxide additions, Am. Ceram. Soc. Bull., 1991, 70, 439-443.

6. Sigl, L.S. and Kleebe, H. J., Core/rim structure of liquid phase-sintered silicon carbide, J. Am. Ceram. Soc., 1993, 76, 773-776.

7. Padture, N. P., In situ-toughened silicon carbide, J. Am. Ceram. Soc.,1994, 77 [2], 519-523.

8. Lee, S. K. and Kim, C. H., Effects of $\alpha$-SiC versus $\beta$-SiC starting powders on microstructure and fracture toughness of SiC sintered with $Al_2O_3$-$Y_2O_3$ additives. J. Am. Ceram. Soc., 1994, 77, 1655-1658.

9. Nagaro, T., Gu, H., Zhan, G. D. and Mitomo, M., Effect of atmosphere on supereplastic deformation behavior in nanocrystalline liquid-phase-sintered silicon carbide with $Al_2O_3$-$Y_2O_3$ additions, J. Mat. Sci., 2002, 37, 4419-4424.

10. Ortiz, A. L., Muñoz-Bernabé, A., Borrero-López, O., Domínguez-Rodríguez, A., Guiberteau, F. and Padture, N. P., Effect of sintering atmosphere on the mechanical properties of liquid-phase-sintered SiC, J. Eur. Ceram. Soc., 2004, 24, 3245-3249 .

11. Schneider, J., Biswas, K., Rixecker, G. and Aldinger F., Microstructural changes in liquid-phase-sintered silicon carbide during creep in an oxidizing environment. J. Am. Ceram. Soc., 2003, 86(3), 501-507.

12. Jou, Z. C. and Virkar, A. V., High temperature creep of SiC densified using a transient liquid phase, J. Mater. Res., 1991, 6 (9), 1945-1949.

13. Gallardo-López, A., Muñoz, A., Martínez-Fernández, J. and Domínguez-Rodríguez, A., High temperature compressive creep of liquid phase sintered silicon carbide, Acta Mater., 1999, 47 (7), 2185-2195.

14 Pujar, V. V., Jensen, R. P. and Padture, N. P., Densification of liquid-phase-sintered silicon carbide, J. Mater. Sci. Lett., 2000, 19 [11], 1011-1014.

15. Ortiz, A. L., Control microestructural de cerámicos avanzados SiC sinterizados con fase líquida $Y_2O_3$ - $Al_2O_3$, PhD Thesis, Universidad de Extremadura, 2002.

16. Ortiz, A. L., Bhatia, T., Padture, N. P. and Pezzotti, G., Microstructural evolution in liquid-phase-sintered SiC: III, Effect of nitrogen-gas sintering atmosphere, J. Am. Ceram. Soc., 2002, 88, 1835-1840.

17. Rouxel, T., Huger, M., Besson, J. L., Rheological properties of Y-Si-Al-O-N glasses elastic moduli, viscosity and creep. J. Mater. Sci., 1992, 27, 279-284.

18. Loehman, R. E., Preparation and properties of oxynitride glasses, J. Non-Cryst. Solids, 1983, 56, 123-134.

19. Sciti, D., Guicciardi, S. and Bellosi, A., Effect of annealing treatments on microstructure and mechanical properties of liquid-phase-sintered silicon carbide, J. Eur. Ceram. Soc., 2001, 21, 621-632.

20. Mah, Tai-II, Keller, K. A., Sambasivan, S. and Kerans, R. J., High-temperature environmental stability of the compounds in the $Al_2O_3$-$Y_2O_3$ system, J. Am. Ceram. Soc., 1997, 80, 874-878.

21. Hon, M. H. and Davis, R. F., Self-diffusion of C-14 in polycrystalline $\beta$-SiC, J. Mater. Sci., 1979, 14, 2411-2421.

22. Hon, M. H., Davis, R. F. and Newbury, D. E., Self-diffusion of Si-30 in polycrystalline $\beta$-SiC. J. Mater. Sci., 1980, 15, 2073-2080.

23. Hon, M. H. and Davis, R. F., Self-diffusion of carbon-14 in high purity and N-doped $\alpha$-SiC single crystals, J. Am. Ceram. Soc., 1980, 63, 546-552.

24. Hon, M. H., Davis, R. F. and Newbury, D. E., Self-diffusion of silicon-30 in $\alpha$-SiC single crystals, J. Mater. Sci., 1981, 16, 2485.

25. Corman, G. S., Creep of $\alpha$-SiC single crystals, J. Am. Ceram. Soc., 1992, 75 (12), 3421.

**Figure captions**

Figure 1: SEM micrographs of samples sintered in different atmospheres and for different times: (a) 3 hours in $N_2$ atmosphere, (b) 1 hour in Ar atmosphere, and (c) 7 hours in Ar atmosphere.

Figure 2: SEM micrographs of samples sintered in an argon atmosphere for: (a) 1 hour, (b) 3 hours, (c) 5 hours, and (d) 7 hours. The dark areas correspond to the grains of SiC and the white areas to the intergranular phase (YAG).

Figure 3: SEM micrographs of samples sintered in a nitrogen atmosphere for: (a) 3 hours and (b) 7 hours.

Figure 4: Distribution of grain size in samples sintered in different atmospheres and for different times: (a) 1 hour in Ar atmosphere, (b) 7 hours in Ar atmosphere, (c) 1 hour in $N_2$ atmosphere, and (d) 7 hours in $N_2$ atmosphere.

Figure 5: Engineering stress-engineering strain curves for uniaxial compression tests at constant speed and a temperature of 1450 °C. Samples sintered for different times and in atmospheres of: (a) Ar and (b) $N_2$.

Figure 6: SEM micrographs of a region close to the surface in deformed samples at 1450 °C. Samples sintered for 1 hour in atmospheres of: (a) Ar and (b) $N_2$.

Figure 7: (a) SEM micrograph of the surface in a deformed sample at 1450 °C. The presence is observed of a surface layer that covers the grains of SiC. (b) EDS microanalysis of the surface layer.

Figure 8: SEM micrographs of deformed samples at 1450 °C. Samples sintered in: (a) Ar atmosphere for 1 hour, (b) Ar atmosphere for 7 hours, and (c) $N_2$ atmosphere for 7 hours.

Table 1: Density of the samples, and comparison with the theoretical density as a function of the atmosphere and sintering time.

| Process time (h) | Measured density (g/cm$^3$) | | Measured density/theoretical density (%) | |
|---|---|---|---|---|
| | Ar atmosphere | N$_2$ atmosphere | Ar atmosphere | N$_2$ atmosphere |
| 1 | 3.27 ± 0.02 | 3.21 ± 0.03 | 98.8 ± 0.6 | 97.0 ± 1.0 |
| 3 | 3.22 ± 0.03 | 3.23 ± 0.03 | 97.3 ± 1.0 | 97.6 ± 1.0 |
| 5 | 3.20 ± 0.02 | 3.22 ± 0.04 | 96.7 ± 0.6 | 97.3 ± 1.3 |
| 7 | 3.14 ± 0.03 | 3.24 ± 0.03 | 94.9 ± 1.0 | 97.9 ± 1.0 |

Table 2: Means and standard deviations of the morphological parameters in the different types of sample, measured before the mechanical compression test. Parameters: equivalent planar diameter d=(4·area/π)$^{1/2}$, form factor F=4π·area/perimeter$^2$, and aspect ratio F$_{asp}$ defined as the ratio between the maximum length and the minimum width taken perpendicular to the length in the planar section of the grains.

| Process atmosphere | Process time (h) | Diameter (μm) | | Form factor | | Aspect ratio | |
|---|---|---|---|---|---|---|---|
| | | $\bar{d}$ | $\sigma_d$ | $\bar{F}$ | $\sigma_F$ | $\bar{F}_{asp}$ | $\sigma_{Fasp}$ |
| Argon | 1 | 0.64 ± 0.02 | 0.29 ± 0.01 | 0.82 ± 0.01 | 0.11 ± 0.01 | 1.62 ± 0.03 | 0.47 ± 0.02 |
| | 3 | 0.97 ± 0.04 | 0.82 ± 0.05 | 0.85 ± 0.01 | 0.09 ± 0.01 | 1.48 ± 0.02 | 0.37 ± 0.01 |
| | 5 | 1.21 ± 0.05 | 0.73 ± 0.03 | 0.85 ± 0.01 | 0.09 ± 0.01 | 1.54 ± 0.02 | 0.39 ± 0.01 |
| | 7 | 1.61 ± 0.05 | 0.99 ± 0.06 | 0.78 ± 0.01 | 0.12 ± 0.01 | 1.87 ± 0.04 | 0.66 ± 0.04 |
| Nitrogen | 1 | 0.50 ± 0.01 | 0.22 ± 0.01 | 0.86 ± 0.01 | 0.09 ± 0.01 | 1.51 ± 0.02 | 0.36 ± 0.01 |
| | 3 | 0.71 ± 0.01 | 0.25 ± 0.01 | 0.88 ± 0.01 | 0.07 ± 0.01 | 1.46 ± 0.02 | 0.26 ± 0.01 |
| | 5 | 0.68 ± 0.02 | 0.26 ± 0.01 | 0.87 ± 0.01 | 0.07 ± 0.01 | 1.50 ± 0.02 | 0.28 ± 0.01 |
| | 7 | 0.68 ± 0.02 | 0.27 ± 0.01 | 0.87 ± 0.01 | 0.08 ± 0.01 | 1.49 ± 0.02 | 0.30 ± 0.01 |

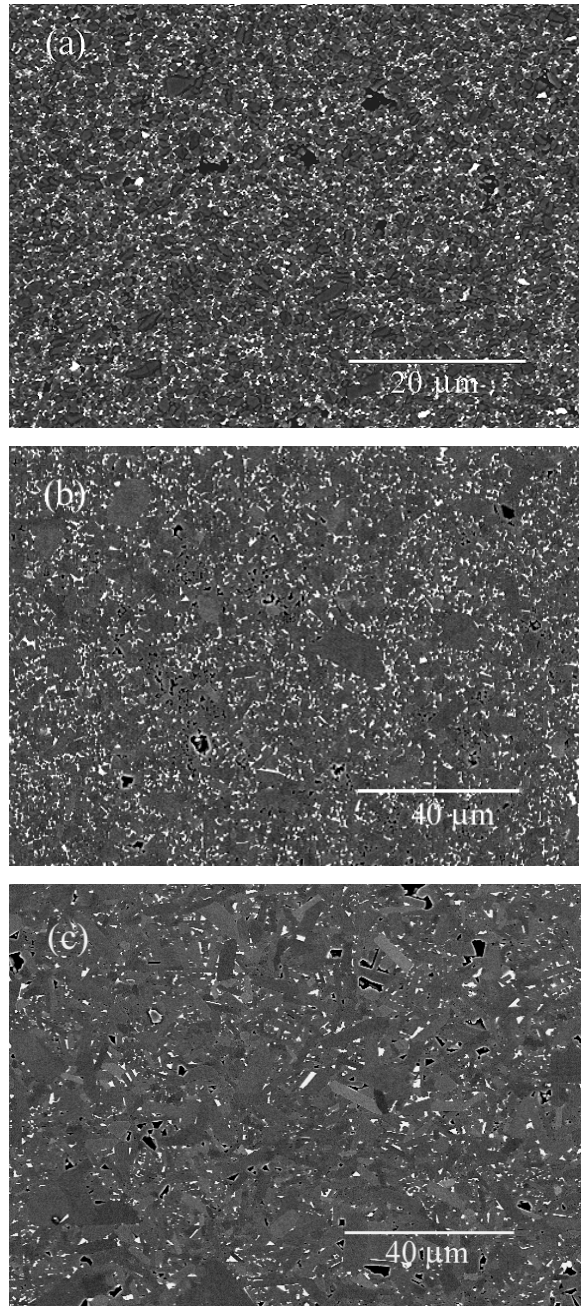

Figure 1

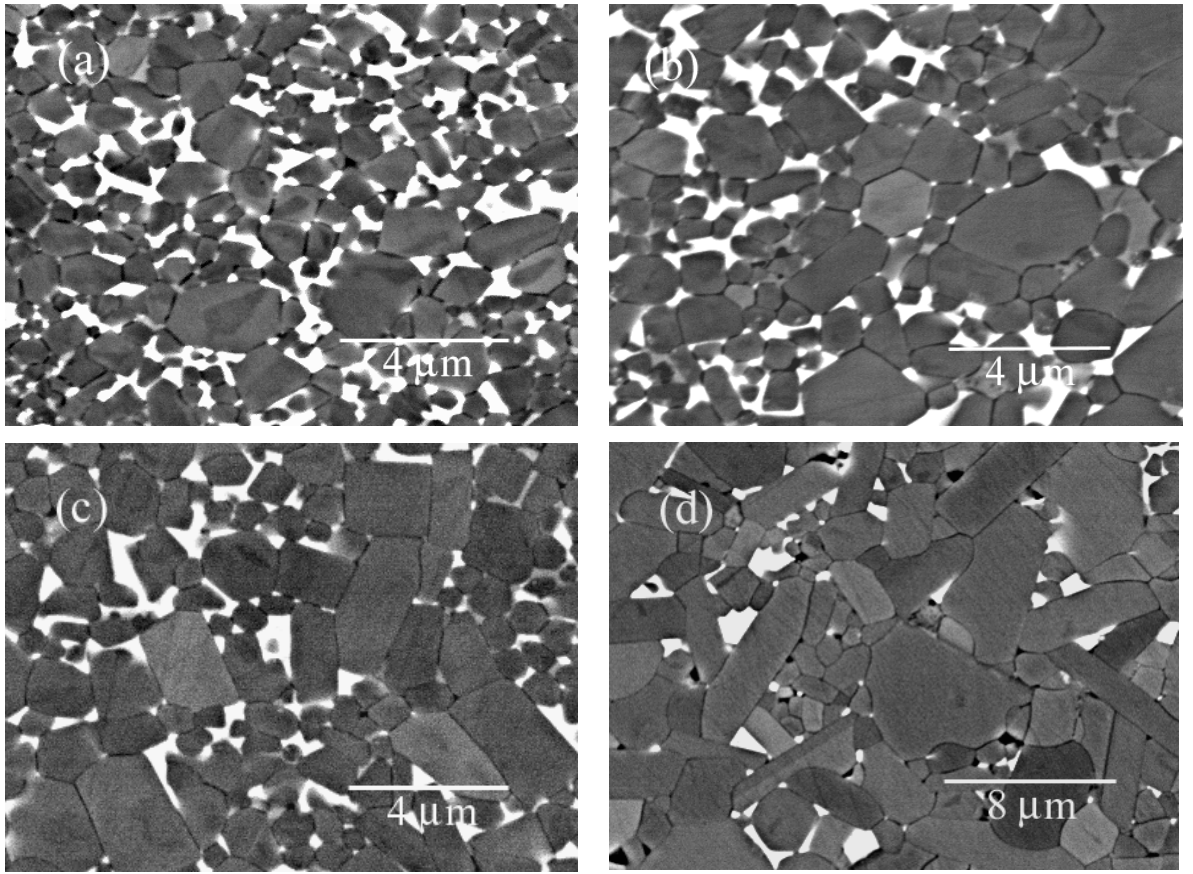

Figure 2

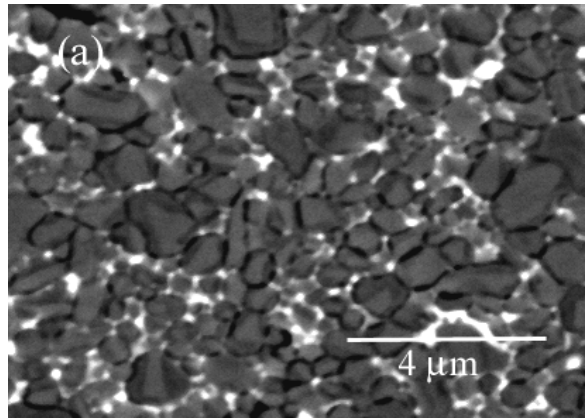

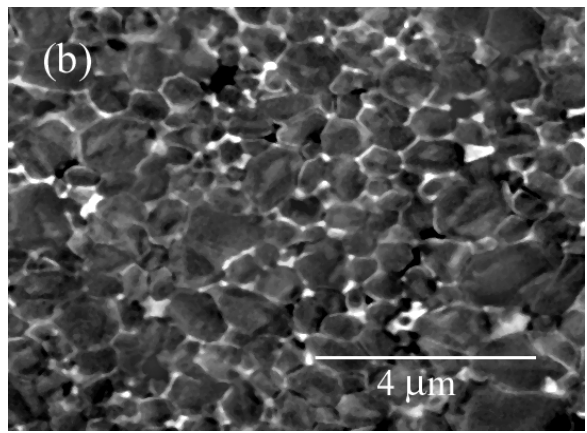

Figure 3

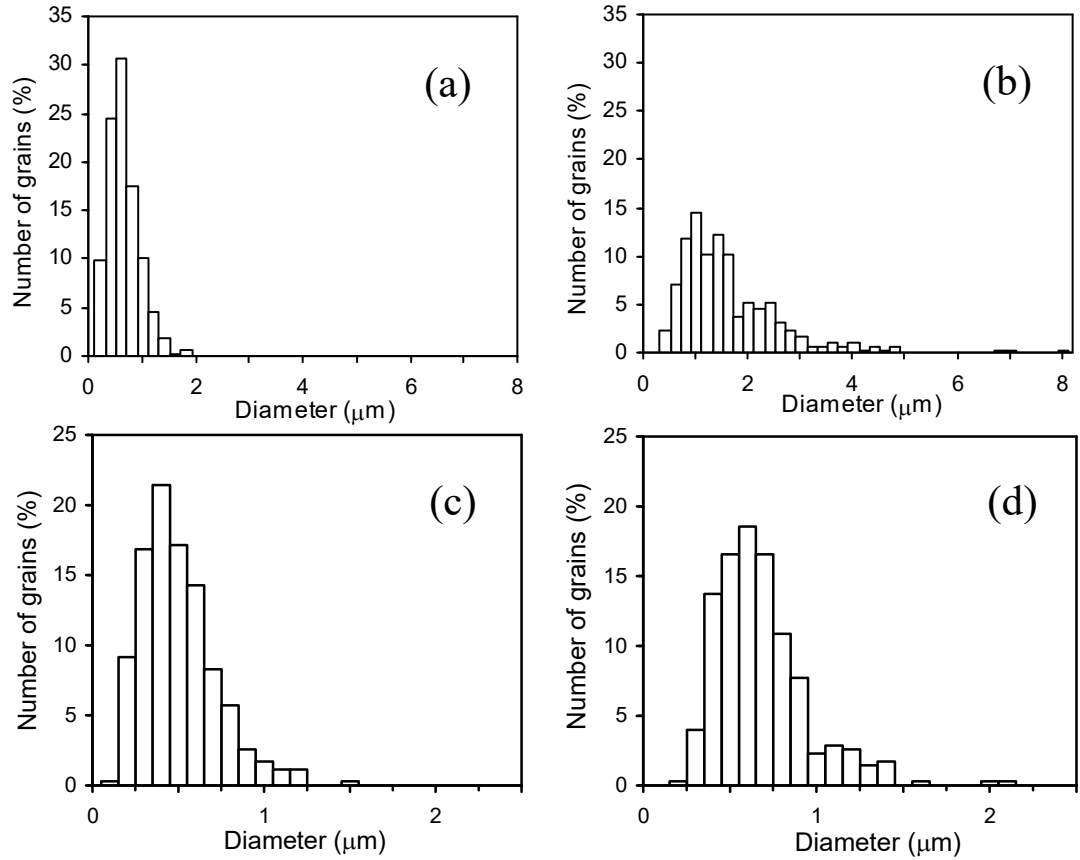

Figure 4

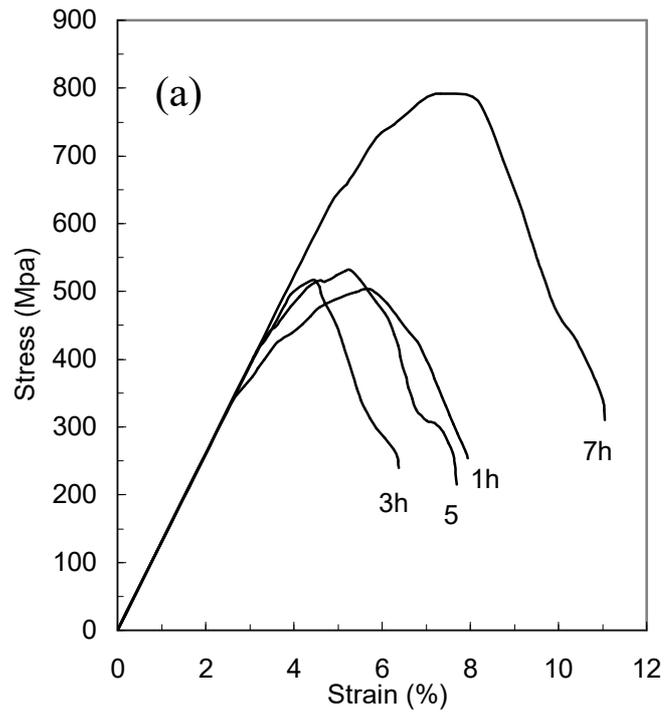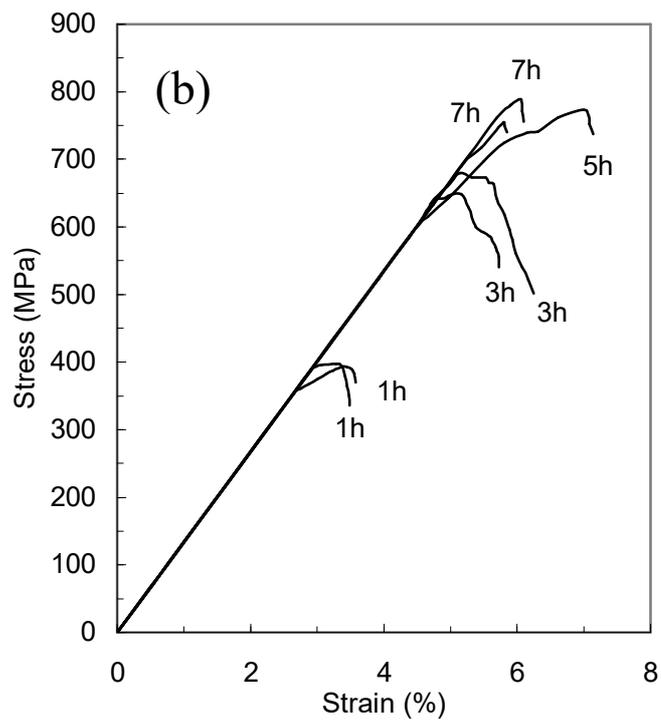

Figure 5

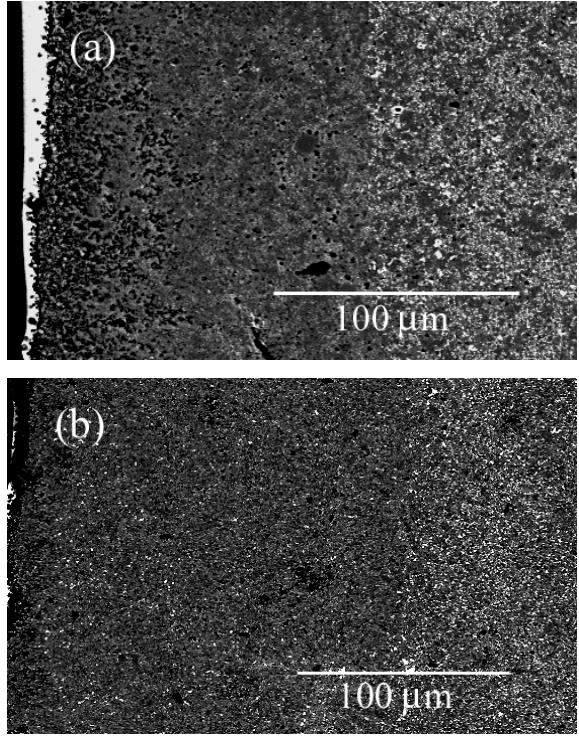

Figure 6

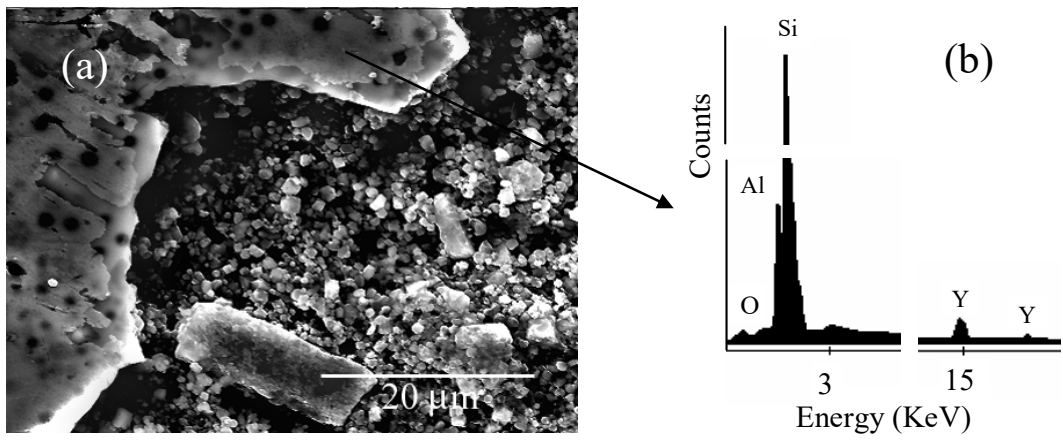

Figure 7

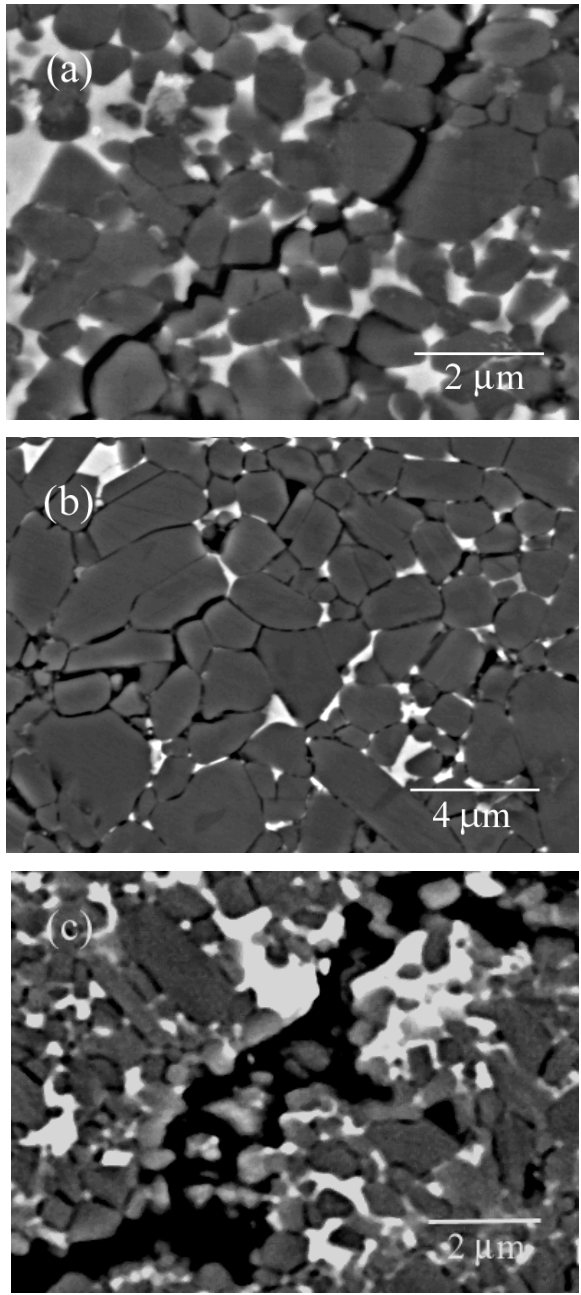

Figure 8